\documentclass[
reprint,
superscriptaddress,
amsmath,amssymb,
aps
prb
]{revtex4-1}
\usepackage{graphicx} 
\usepackage{breqn}
\usepackage{braket}
\usepackage{amsmath}
\usepackage{float}
\usepackage{graphicx, caption,subcaption}
\usepackage{float}
\usepackage{gensymb}
\usepackage{subfig}
\usepackage{lipsum}
\usepackage{dcolumn}
\usepackage{bm}
\usepackage{xcolor} 

\usepackage [english]{babel}
\usepackage [autostyle, english = american]{csquotes}
\MakeOuterQuote{"}
\begin{document}

\title{Increased Gilbert damping in Yttrium Iron Garnet by low temperature vacuum annealing}

\author{D. M. Cheshire}
\affiliation{School of Physics$,$ Engineering and Technology$,$ University of York$,$ York$,$ YO10 5DD$,$ UK}

\author{D. Backes}
\affiliation{Diamond Light Source$,$ Harwell Science and Innovation Campus$,$ Chilton$,$ Didcot$,$ OX11 ODE$,$ UK}

\author{L. Ishibe Veiga}
\affiliation{Diamond Light Source$,$ Harwell Science and Innovation Campus$,$ Chilton$,$ Didcot$,$ OX11 ODE$,$ UK}

\author{S. S. Dhesi}
\affiliation{Diamond Light Source$,$ Harwell Science and Innovation Campus$,$ Chilton$,$ Didcot$,$ OX11 ODE$,$ UK}

\author{S. A. Cavill}
\altaffiliation{stuart.cavill@york.ac.uk}
\affiliation{School of Physics$,$ Engineering and Technology$,$ University of York$,$ York$,$ YO10 5DD$,$ UK}

\begin{abstract}
The effect of thermal surface cleaning on the Gilbert damping ($\alpha$) of yttrium iron garnet (YIG), before capping with a metallic layer, has been investigated. Our results show that $\alpha$ is strongly effected by relatively mild annealing conditions ($T$ = 300 $\degree$ C) when performed in a vacuum. This increase needs to be taken into account when obtaining the spin mixing conductance from spin pumping measurements.
We measure an increase in $\alpha$ by a factor of $\times$8 when the YIG is vacuum annealed at 300 $\degree$ C. No such changes in $\alpha$ are observed when annealed at the same temperature in 1$\times$10$^{-1}$ mbar of oxygen.
\end{abstract}
\maketitle
\section{Introduction}
Research in the sub-field of magnon spintronics, where magnons - the quanta of spin waves - are used to carry and process information has gathered pace in recent years \cite{Chumak2022, Flebus2024,Chumak2015453}. Magnons offer several advantages over conventional spintronics, particularly in the development of insulator-based devices with reduced energy consumption. In these devices, the propagation of pure spin currents in magnetic insulators, where dissipationless transport of the spin has been predicted \cite{Takei2014}, eliminates parasitic Joule heating. One magnetic insulator in particular, yttrium iron garnet (YIG), is considered the most prominent material in this field,
being widely used in spin transport experiments \cite{Cornelissen20151022} due to its
exceptionally low damping, even in thin films \cite{Onbasli2014}. In order to combine magnon spintronics and charge based electronics, the latter crucial for input - output functionality, it is necessary to create efficient converters that transform magnon spin currents into conventional charge currents. Typically, spin pumping, the generation of spin accumulation a metal by the magnetization precession of an adjacent ferromagnetic, and the inverse spin Hall effect (ISHE), that transforms a pure spin current into a charge current, are used for the input and output conversion processes respectively. According to spin-pumping theory \cite{Tserkovnyak20022244031}, the spin current sunk
by a layer adjacent to a ferromagnet (FM) leads to an increase in the Gilbert damping such that \(\alpha=\alpha_{intrinsic}+\alpha_{sp}\), with 
\begin{equation}
\alpha_{sp}=\frac{g\mu_B}{4\pi M_s}g_{eff}^{\uparrow\downarrow}\frac{1}{d}
\end{equation}
where $M_s$ is the saturation magnetization of the FM, $d$ the
FM layer thickness, and $g_{eff}^{\uparrow\downarrow}$ is the effective spin-mixing
conductance that essentially governs the transfer of spin momentum across an interface.
Measuring the change in the Gilbert damping of the YIG due to spin pumping into an adjacent metallic layer therefore allows a convenient experimental determination of $g_{eff}^{\uparrow\downarrow}$.\\ Previous studies of the spin mixing conductance at a YIG/Pt interface demonstrated an enhancement by more than two orders of magnitude by surface 
treatments, such as Piranha etch and heating the sample at 500 \degree C in vacuum, prior to Pt deposition \cite{Jungfleisch2013}. The authors considered that the intrinsic Gilbert damping is the same for all samples, irrespective of treatment method, so that changes in damping is solely related to changes in $g_{eff}^{\uparrow\downarrow}$.
In this paper, we report increases in the Gilbert damping in YIG thin films, due to either being subjected to vacuum thermal annealing, or the subsequent deposition of a thin (2 nm) weak spin-orbit coupled aluminium capping layer. The Gilbert damping is observed to increase significantly for YIG, vacuum annealed at 300 \degree C for as little as 15 minutes, whereas for YIG annealed at the same temperature in a 1$\times 10^{-1}$ mbar partial pressure of oxygen, $\alpha$ is unaffected. In addition, we notice that the deposition of Al on the YIG, irrespective of film heat treatment, also increases $\alpha$.  We argue that such increases need to be considered when forming spin pumping heterostructures as often the YIG films are annealed in vacuum to remove physisorbed gas species (O$_2$, air) from the surface prior to heavy metal deposition. These increases in $\alpha$ may then be incorrectly attributed to $\alpha_{sp}$, leading to errors in $g_{eff}^{\uparrow\downarrow}$. 
We suggest that the main driver for the increase in $\alpha$ is the reduction of Fe$^{3+}$ to Fe$^{2+}$, consistent with vacuum annealing and / or a displacement reaction between Fe oxide and Al.  Changes in the Fe $L_{2,3}$ X-ray absorption spectra, with annealing conditions and capping layer material, support this hypothesis.

\section{Methods}
Crystalline YIG films were grown via pulsed laser deposition (PLD) on (111)-oriented  Gadolinium Gallium Garnet (GGG) substrates. Following \cite{Hauser2016, Cheshire2024}, an amorphous YIG layer was deposited at room temperature, in a partial oxygen pressure of 2.5$\times 10^{-3}$ mbar, by ablating a stoichiometric polycrystalline YIG target. Ablation was performed using a frequency quadrupled Nd:YAG laser ($\lambda$ = 266 nm) with a fluence of approximately 1 Jcm$^{-2}$, at a repetition frequency of 10 Hz. The sample was rotated at 6 RPM during deposition. Following deposition, the amorphous YIG samples were ex-situ annealed in atmospheric conditions, at 850 \degree C for 3 hours, to recrystallise the YIG. This technique was used to produce four YIG films with nominal thicknesses of (65 $\pm$ 3) nm, for FMR measurements.\\
Post FMR, the four YIG films were then subjected to surface treatments, as follows. Two films were placed into the PLD chamber, in vacuum (at a base pressure of 5x10$^{-8}$ mbar) and annealed at 150 \degree C and 300 \degree C respectively, for 15 minutes. In-situ annealing was performed with a pulsed 10.6 µm CO$_2$ laser, directed onto the back of the samples. The temperature of each film was ramped at approximately 4 \degree C/min, by varying the laser’s duty cycle. The third film was subjected to annealing at 300\degree C following the same procedure, but instead in an oxygen partial pressure of 1$\times 10^{-1}$ mbar. The fourth film was left untreated as a control.   FMR measurements were then repeated on the treated films to observe changes to the magnetic damping. The thermal annealing treatments were repeated on 2 further sets of recrystallized YIG films to ensure repeatability.
An additional 2 nm aluminium capping layer was deposited on three YIG films by sputtering a pure Al target at room temperature. A final set of FMR measurements were performed on the aluminium capped YIG films. The aluminium capping was also performed to allow total electron yield (TEY) detection for XAS/X-ray magnetic circular dichroism (XMCD) spectroscopy measurements. The fourth YIG film was grown and capped with a 2nm platinum capping layer as a control. The Pt layer was deposited at room-temperature using PLD, ablating a pure Pt target in vacuum (base pressure of 5$\times 10^{-8}$ mbar). \\
The Gilbert damping of the YIG was measured using in-plane FMR spectroscopy by mounting the samples face down onto a 50$\Omega$ coplanar waveguide (CPW), connected to a two-port vector network analyser (VNA) and centred between the poles of a 2D vector magnet. Details of the system can be found in \cite{Love2021}. The microwave transmission S-wave parameter, S$_{12}$, was measured over a frequency range of 0.001-15 GHz. The applied magnetic field ranged from 0-4 kOe, with a field step of 0.5 Oe, producing a 2D frequency-field resonance map. All FMR measurements were performed at an RF power of +7 dBm. Field linescans at constant frequency were extracted from the FMR frequency-field maps and fitted using an asymmetric Lorentzian function \cite{Love2023} to determine the corresponding resonance field (H$_r$) and linewidth ($\Delta$H). \\
Fe $L_{2,3}$ x-ray absorption spectroscopy (XAS) and XMCD measurements were performed on beamline I06 at the Diamond Light Source \cite{Dhesi2010311}. The XAS was measured in normal incidence by total-electron yield (TEY) using the sample drain current.
\\
\section{Results}
Resonant frequencies were extracted from the frequency-field maps for each applied field and are plotted in Fig.1 for the YIG films prepared under different conditions. The FMR resonance frequency as a function of field is fit to the easy-axis in-plane Kittel equation for a (111) orientated film, given by

\begin{equation}
\omega=2\pi\gamma\sqrt{\big(H\big(H+4\pi M_{eff}\big)\big)}
\end{equation}
with the gyromagnetic ratio \(\gamma=g\mu_B / \hbar\), and the effective magnetisation \(M_{eff}=4\pi M_s-K/M_s\) where $M_s$  is the saturation magnetisation of the YIG, and K is the cubic magnetocrystalline anisotropy constant (-6100 erg/cc for YIG) \cite{Hansen1974}. 
\begin{figure}[h]
    \centering
    \includegraphics[width=1\linewidth]{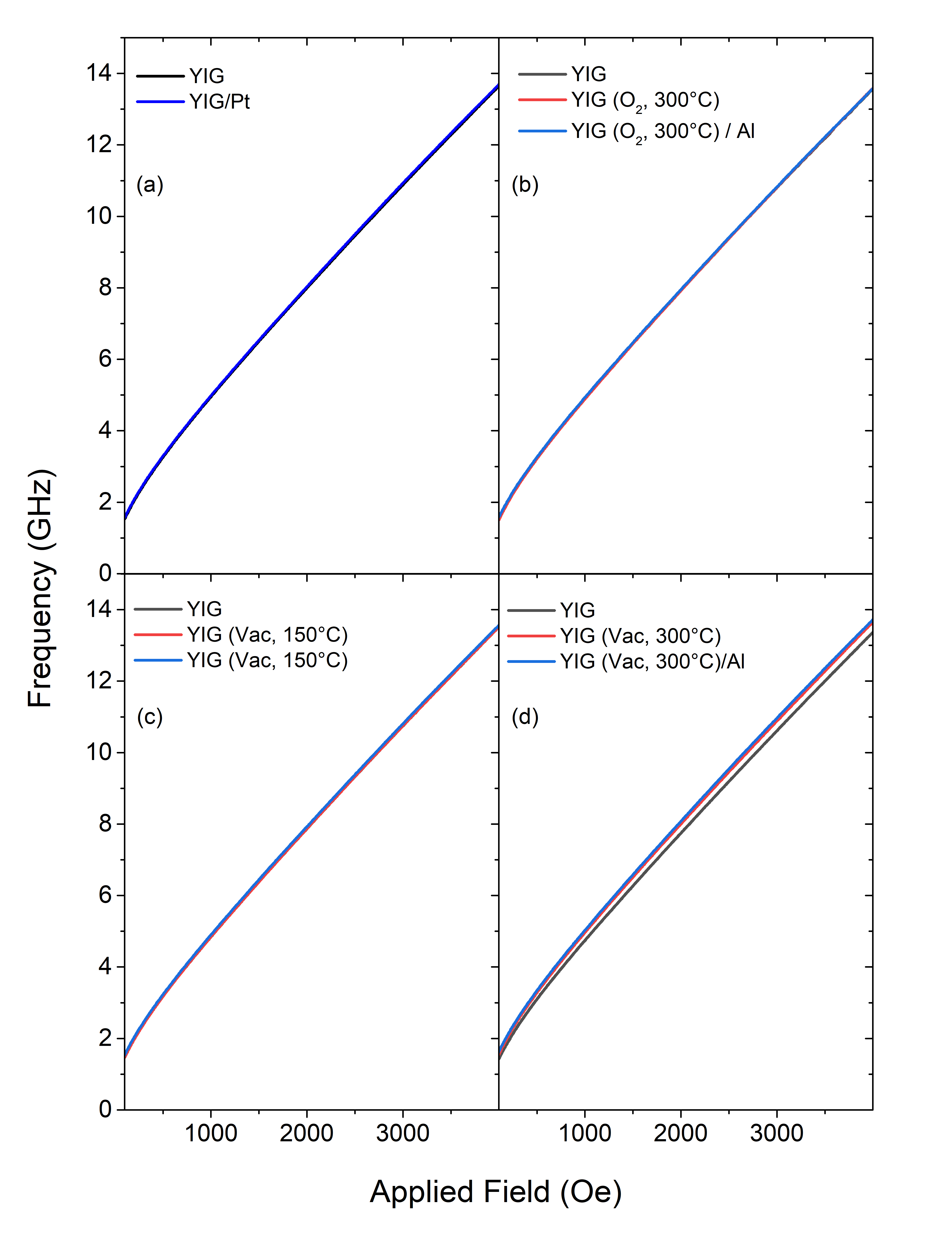}
    \caption{Kittel curves (frequency vs field) for the (black) as-grown,
  (red) annealed YIG films, and (blue) annealed YIG films after capping with 2nm metal layer.}
    \label{fig:enter-label}
\end{figure}

For three of the YIG films – untreated, oxygen annealed at 300\degree C, vacuum annealed at 150 \degree C - the FMR resonant field (measured at 10 GHz) of all three stages differ by less than 10 Oe, with the "Kittel" curves almost lying on top of each other. However, for YIG vacuum annealed at 300 \degree C (Figure 1d), a more significant decrease in resonant field ($H_r$) of 95 Oe is observed, with an additional decrease of 18 Oe following aluminium capping. Nevertheless, "Kittel" curve fits for all four films produce values for $M_s$ that agree within error with independent measurements of the untreated YIG using vibrating sample magnetometry, (131 $\pm$ 5) emu/cc. The three films with similar Kittel curves also produce a similar $g$-factor of (2.02 $\pm$ 0.01) across all stages of these three samples, close to the expected value of 2 for a d$^5$ ion \cite{Cheshire2022}. However, the  300 \degree C vacuum-annealed produces a larger g-factor of (2.050 $\pm$ 0.005). 
\begin{figure}[h]
    \centering
    \includegraphics[width=1\linewidth]{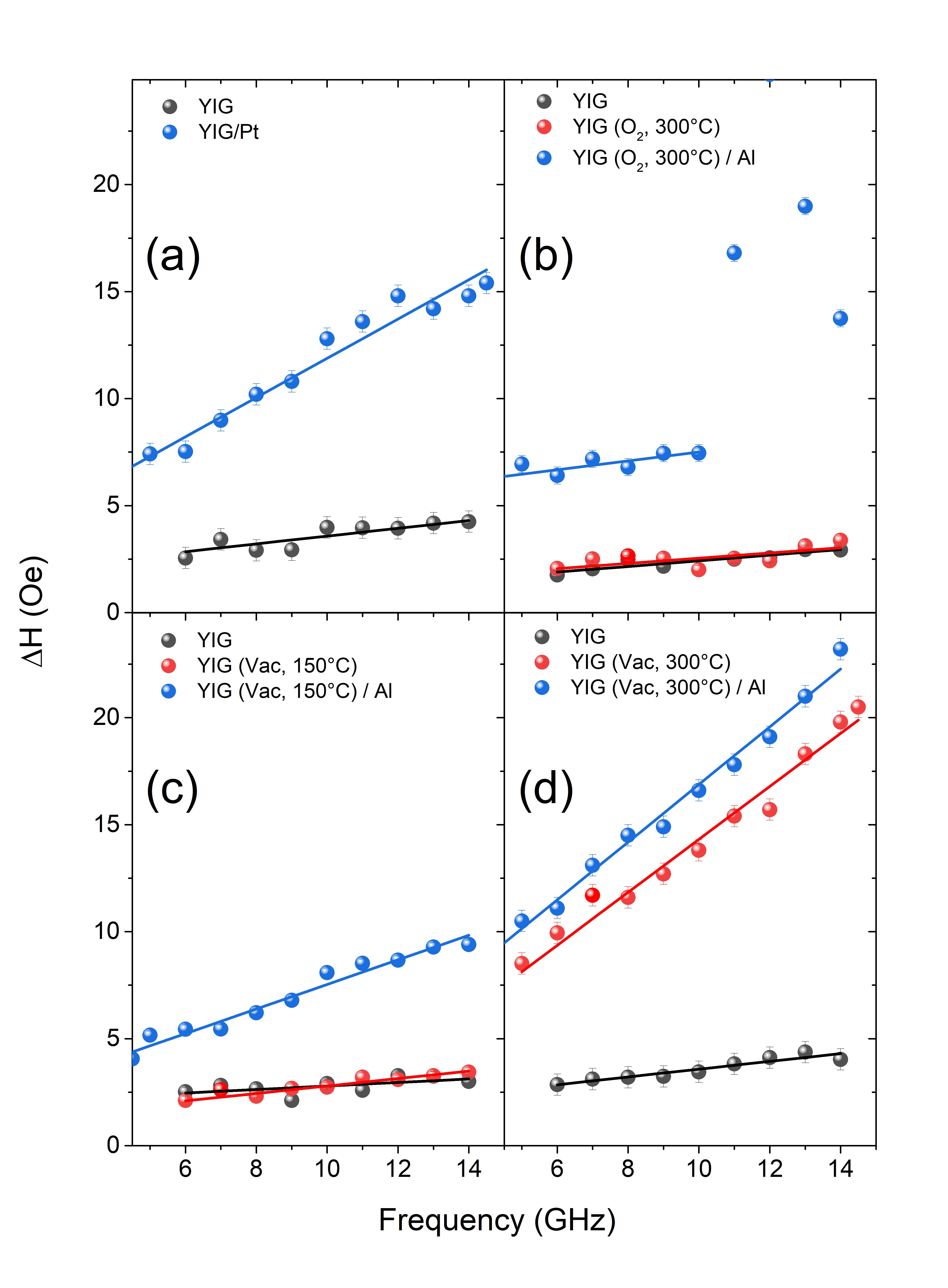}
    \caption{FMR linewidth ($\Delta$H) vs frequency for the (black) as-grown,
  (red) annealed YIG films, and (blue) annealed YIG films after capping with 2nm metal layer.}
    \label{fig:enter-label}
\end{figure}
\\In samples where two-magnon scattering is negligible, the magnetic linewidth ($\Delta$H) of the FMR resonance follows a linear dependence with microwave frequency, such that \cite{Parkes2013}

\begin{equation}
\Delta H(f)=\Delta H_0 +\frac{2\pi\alpha f}{\gamma}
\end{equation}

where $\Delta H$ is the half width at half maximum (HWHM) of the resonance, $\alpha$ is the intrinsic Gilbert damping, $\gamma$ the gyromagnetic ratio and $\Delta H_0$ is the frequency independent extrinsic damping. 
The experimental data for magnetic linewidth as a function of microwave frequency are shown in Figure 2.

Fitting Eqn (3) to the data allows for the extraction of the Gilbert damping ($\alpha$) and frequency independent extrinsic damping ($\Delta H_0$), for each of the treated YIG conditions. $\Delta H$ vs $f$ is linear for all samples (prior to Al deposition), indicating that two-magnon scattering is not significant in our samples, even those annealed in-vacuum. 
We observe no significant increases in the Gilbert damping that arise from in-vacuum annealing of the YIG at 150 \degree C. However, in-vacuum annealing at 300 \degree C – even for only 15 minutes - dramatically raises the Gilbert damping, by almost an order of magnitude upon re-measurement of the FMR: from (5.0 $\pm$ 0.6)$\times 10^{-4}$ to (38 ± 1)$\times 10^{-4}$. In contrast, for annealing performed at 300 \degree C in 1$\times 10^{-1}$ mbar of pure oxygen, the Gilbert damping was found to be, within error, insignificantly changed. This is an important observation and highlights the sensitivity of YIG FMR to preparation conditions, such as thermal cleaning of physisorbed species, that may be used when depositing multiple layers for magnonic devices.

\begin{figure}[h]
    \centering
    \includegraphics[width=1\linewidth]{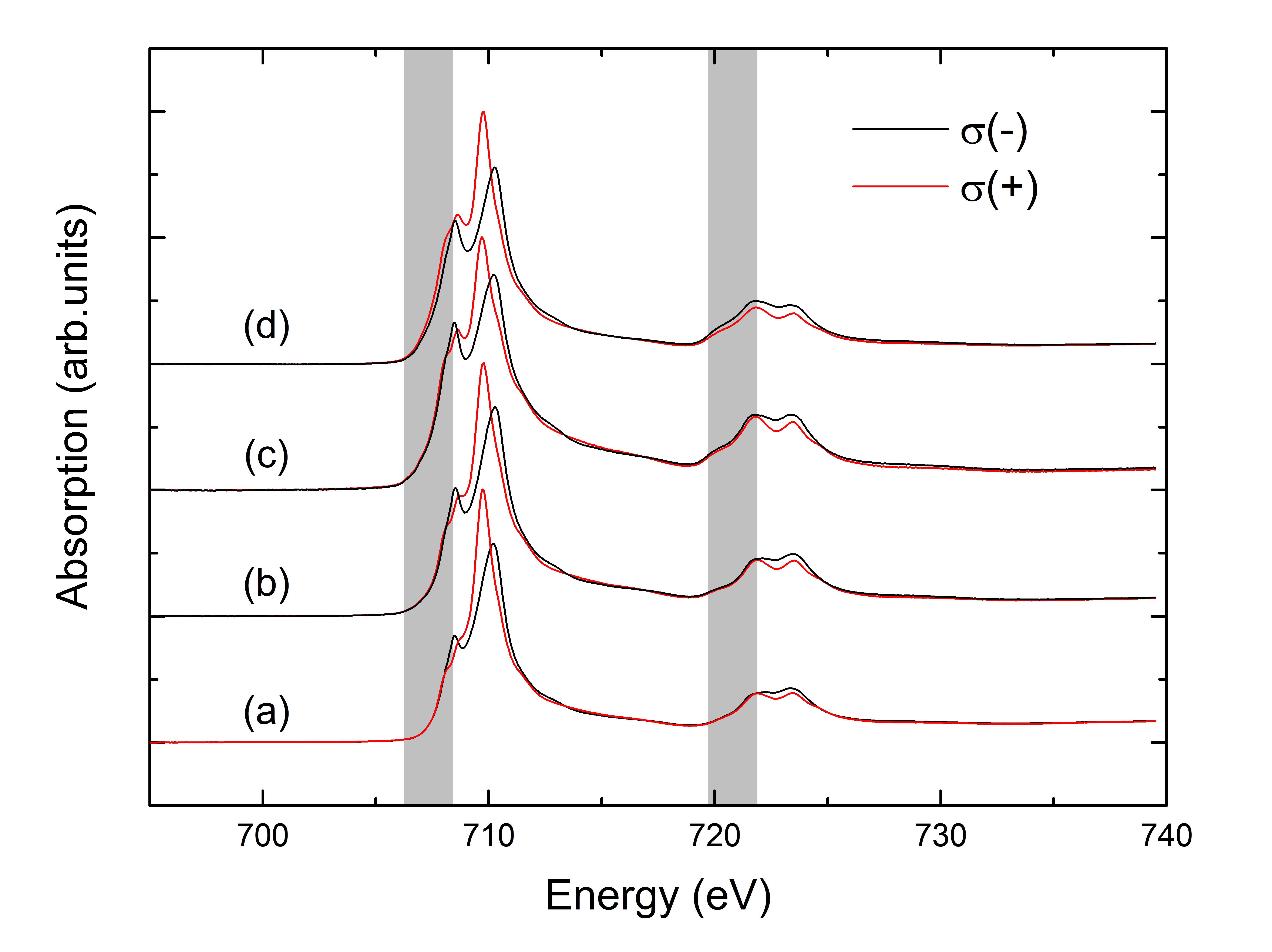}
    \caption{Circular polarization-dependent XAS at the Fe $L_{2,3}$ edges for the (a) YIG/Pt, (b) YIG(vac, 150\degree C)/Al, (c) YIG($O_2$, 300\degree C)/Al and (d) YIG(vac, 300\degree C)/Al samples}
    \label{fig:enter-label}
\end{figure}
No obvious relationship between annealing treatment and extrinsic damping was inferred from the FMR data. In all three annealing treatments, the changes in extrinsic damping are either within error, or insignificantly small, with $\Delta H_0$ of (1.5 $\pm$  0.5) Oe for all cases. This suggests that the annealing temperatures used, despite being high enough to affect the Gilbert damping when performed in vacuum, does not significantly impact the YIG/GGG interface or promote the formation of defects in the bulk of the YIG film.
Interestingly, the addition of an aluminium layer causes a notable increase in Gilbert damping for all the YIG films. At first consideration, this is not expected as aluminium (Z = 13) is non-magnetic with a weak spin-orbit coupling and therefore would not be a natural choice for a spin-sink. The low layer thickness (2 nm) and natural oxidation, after removal from the sputtering chamber, precludes significant eddy current damping as the cause. In addition, we rule out radiative damping \cite{Qaid2017, Schoen2015} as significant, as we obtain $\alpha^{rad}$ = 2 $\times$ 10$^{-5}$ from Eqn. 6 in \cite{Schoen2015} for our geometry.  However, an unexpected second FMR resonance mode was observed in the oxygen-annealed YIG film following aluminium deposition. The presence of this second mode restricted the frequency range across which the HWHM linewidths of the Kittel mode could be accurately determined, to below 10 GHz as reflected in Figure 2(b), as above this frequency the two modes significantly overlapped broadening the resonance. It is not clear the cause of this second resonance, but we speculate that it may be due to an inhomogeneous Al layer or a slight increases in the roughness of the YIG film at the YIG/Al interface, induced by the sputtering process. However, it is clear from the slope of $\Delta H$ vs frequency that there is little change in $\alpha$ after annealing at  300 \degree C in oxygen.

\begin{figure}[h]
    \centering
    \includegraphics[width=1\linewidth]{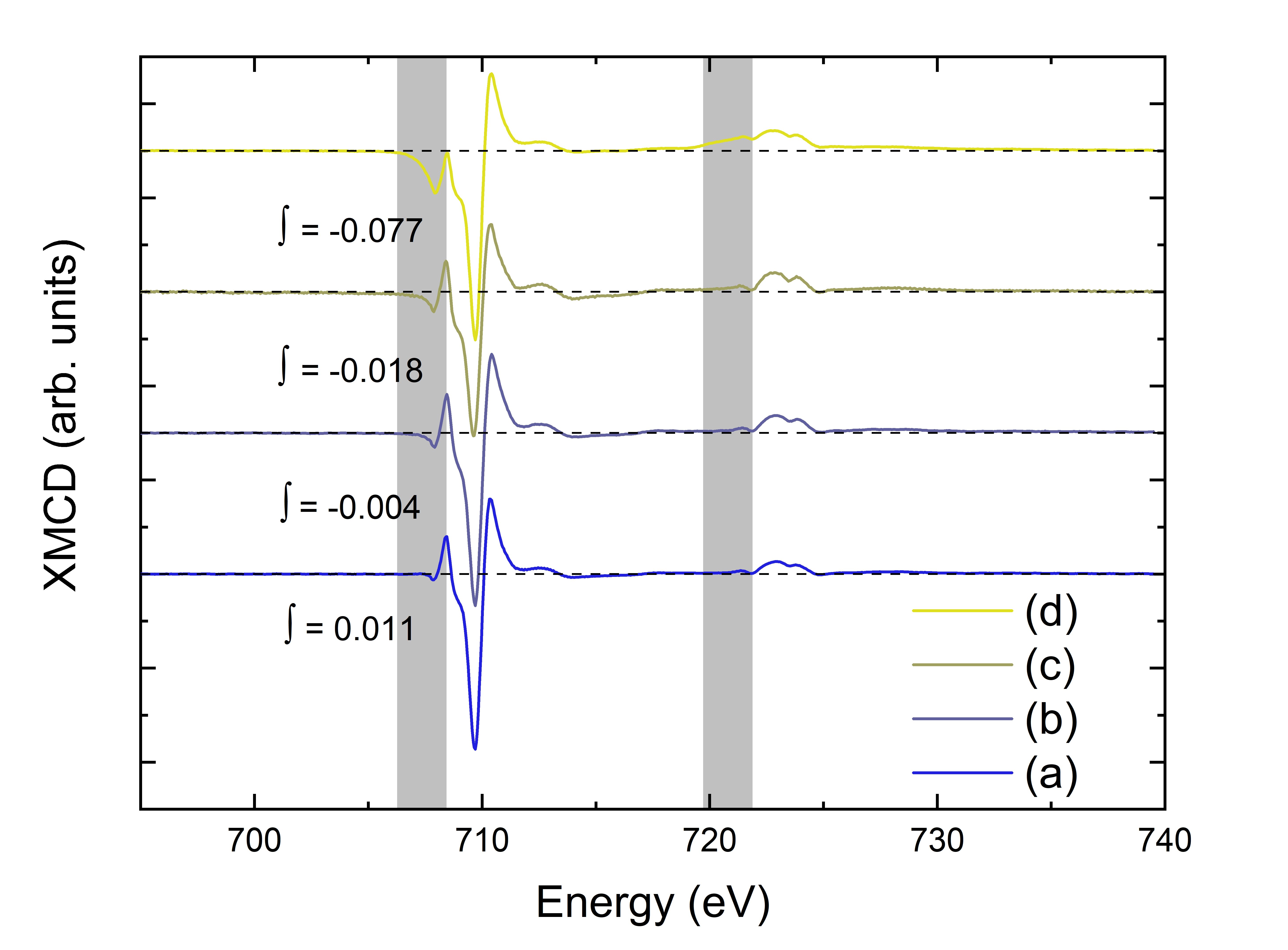}
    \caption{XMCD spectra at the Fe $L_{2,3}$ edges for: (a) YIG/Pt, (b) YIG(vac, 150\degree C)/Al, (c) YIG($O_2$, 300\degree C)/Al and (d) YIG(vac, 300\degree C)/Al samples.}
    \label{fig:enter-label}
\end{figure}
To gain a detailed understanding of the annealed YIG, XAS/XMCD measurements were performed at the Fe $L_{2,3}$ absorption edges in order to determine the valency, coordination, and magnetic properties of the Fe
cations. In L-edge XAS, electrons are excited from a 2p
core level to the unoccupied 3d valence states of the element
of interest by circularly polarized x-rays at the resonance
energy of the transition. The difference in absorption for opposite chirality of the polarizations (XMCD) gives a direct
and element-specific measurement of the projection of the
3d magnetic moment along the x-ray helicity vector \cite{vanderLaan2014X-rayMagnetism}.
At the $L_{2,3}$ absorption edges of 3d transition metals, the probing depth
for TEY detection is approximately 6 nm. All measurements
were performed at normal incidence, which reduces the effect
of self-absorption on the spectra \cite{Nakajima1999}, in a 6 T field applied
collinear to the x-ray helicity vector.
Figure 3 shows XAS data for the (a) YIG/Pt, (b) YIG(vac, 150\degree C)/Al, (c) YIG($O_2$, 300\degree C)/Al and (d) YIG(vac, 300\degree C)/Al samples at the Fe $L_{2,3}$
edges. The XAS shows
multiplet structure typical of YIG \cite{Cheshire2022}. Two main peaks, one negative and one positive, are present in the XMCD spectra at the $L_3$ edge, identical to that found in other XMCD studies of YIG \cite{Cheshire2022}. These peaks correspond to contributions from Fe$^{3+}$ in octahedral ($O_h$) and Fe$^{3+}$ in tetrahedral ($T_d$) sites. For the ferrimagnetic garnet structure, spins located on the $T_d$ sites are aligned antiparallel to the spins on the $O_h$ sites in a ratio of 3:2. Hence, the negative peak in the XMCD at the $L_3$ edge is due to tetrahedrally coordinated Fe$^{3+}$. However, on annealing in an increasingly reducing environment we notice a change in both the XAS and XMCD close to the pre-edge region of both the $L_3$  and $L_2$ edges, as shown in the grey boxed regions of Figs.3,4. The clear increase in spectral weight that occurs towards lower photon energies is indicative of an increase in Fe$^{2+}$, consistent with annealing in a reducing vacuum environment and  / or a displacement reaction between iron oxide and metallic aluminium. To further demonstrate this, we integrate the XMCD in the pre-edge region of the $L_3$ edge (grey box). For the Pt capped sample the integral is slightly positive, shown in Fig. 4, but becomes increasingly negative as the samples are progressively reduced. This indicates an increasing amount of Fe$^{2+}$, aligned parallel to the $T_d$ Fe$^{3+}$ in the near surface region. As Fe$^{2+}$ has a larger orbital moment compared to Fe$^{3+}$, any increase in the former would provide an additional relaxation mechanism to the lattice, thereby enhancing the Gilbert damping, consistent with the FMR results.

\section{Conclusion}
In conclusion, we demonstrate that moderate vacuum annealing of YIG or the addition of an aluminium capping layer significantly increases the Gilbert damping. For the former case, this increase needs to be taken into account when obtaining the spin mixing conductance from spin pumping measurements. We suggest that the main driver for the increase in $\alpha$ is the reduction of Fe$^{3+}$ to Fe$^{2+}$, as demonstrated by soft X-ray magnetic spectroscopy.

\section{Acknowledgments}
The authors acknowledge Diamond Light Source for time on Beamline I06 under proposal MM32492.

\medskip

\bibliographystyle{unsrt}
\bibliography{YIG3}
\end{document}